\documentclass{svjour3}     
\usepackage{graphicx}

\usepackage{amssymb}
\usepackage{amsmath}

\usepackage{amsfonts}
\usepackage{array}
\usepackage{multirow}

\usepackage{natbib}

\usepackage{grffile}

\DeclareGraphicsExtensions{.png,.pdf}

\usepackage{subfigure}
\graphicspath{{./figures/}}

 \journalname{Celestial Mechanics and Dynamical Astronomy}

\usepackage{color}

\begin{document}




\title{An Elementary Solution To Lambert's Problem}


\author{Robert W. Easton           \and
        Rodney L. Anderson         \and
        Martin W. Lo             
}


\institute{
           Robert W. Easton \at
           Applied Mathematics,
           University of Colorado at Boulder,
           Boulder, CO 80309 \\
              \email{Robert.Easton@colorado.edu} 
           \and
Rodney L. Anderson \at
           Jet Propulsion Laboratory,
           California Institute of Technology,
           4800 Oak Grove Drive,
           M/S 301-121,
           Pasadena, CA 91109\\
              \email{Rodney.L.Anderson@jpl.nasa.gov}           
           \and
           Martin W. Lo \at
           Jet Propulsion Laboratory,
           California Institute of Technology,
           4800 Oak Grove Drive,
           M/S 301-121,
           Pasadena, CA 91109 \\
              \email{Martin.W.Lo@jpl.nasa.gov} \\
\copyright \hspace{1pt} 2021 California Institute of Technology. Government sponsorship acknowledged.
}

\date{Received: date / Accepted: date}

\titlerunning{An Elementary...}

\maketitle


\begin{abstract}
A fundamental problem in spacecraft mission design is to find a
\textcolor{black}{free-flight} path from one place to another with a given
transfer time. This problem for paths in a central force field is known as
Lambert's problem.  Although this is an old problem, we take a new approach.
Given two points in the plane, we produce the conic parameters $(p,e,\omega)$
for all conic paths between these points. For a given central force
gravitational parameter, the travel time between the launch and destination
points is computed along with the initial and final velocities for each
transfer conic. For a given travel time, we calculate the parameters for a
transfer conic having that travel time.
\end{abstract}

\section{Introduction} 

Lagrange showed in 1778 that the time required to traverse an elliptic arc
between specified endpoints for Kepler's problem depends only on the semi-major
axis of the ellipse, the distance between the two points, and the sum of their
distances from the focus of the ellipse. 
A wide range of methods for solving Lambert's problem have subsequently been
proposed over the years with various areas of applicability for different
problems.  These methods often focus on the use of various parameters such as F
\& G series \citep{Bate:1971}, universal variables \citep{Vallado:2013}, and 
\textcolor{black}{semimajor axis}
\citep{Prussing:2013} just to mention a few.  Other standard solutions to
Lambert's problem can be found in \cite{Pollard:1966}, \cite{Battin:1999}, and
\cite{Gooding:1990}.  A more complete description of various Lambert problem
solutions may be found in \cite{Russell:2019}.  Many of these approaches focus
on the proper selection of variables to use in combination with the
time-of-flight to optimize subsequent convergence on a desired time-of-flight
using various iteration methods.

\textcolor{black}{
The textbook solution to Lambert's problem typically
derives the transfer time using Kepler's equation and the eccentric anomalies
of the endpoints.}
To calculate travel times between points on a transfer conic arc we use
conservation of angular momentum and numerical integration. This avoids using
Kepler's equation and eccentric anomalies. 
In our approach, we first focus on determining the set of all elliptical orbit
solutions that connect two points.  We use a geometric approach based on
standard orbital elements that eliminates the need for using various
intermediate parameters that may not always have an obvious physical meaning.
In particular, \textcolor{black}{the conic parameters $p$ and $e$  (the scale parameter
and eccentricity)} form a triangle in the
parameter plane for the elliptical transfer orbit case, which is the basis of
our approach.

In the final section, the results presented here are used to construct a
sequence of conic arcs between a sequence of patch points \textcolor{black}{ as
an application}. The method \textcolor{black}{is general and} can be extended to
construct conic arcs between patch points in three dimensions. The method can
be used as a preliminary mission design tool producing a sequence of conic arcs
between a sequence of patch points \textcolor{black}{which could include
maneuvers or flybys}.

\section{A Geometry Problem}
\begin{figure}[ht!]
\centering
\includegraphics[width=0.7\textwidth]{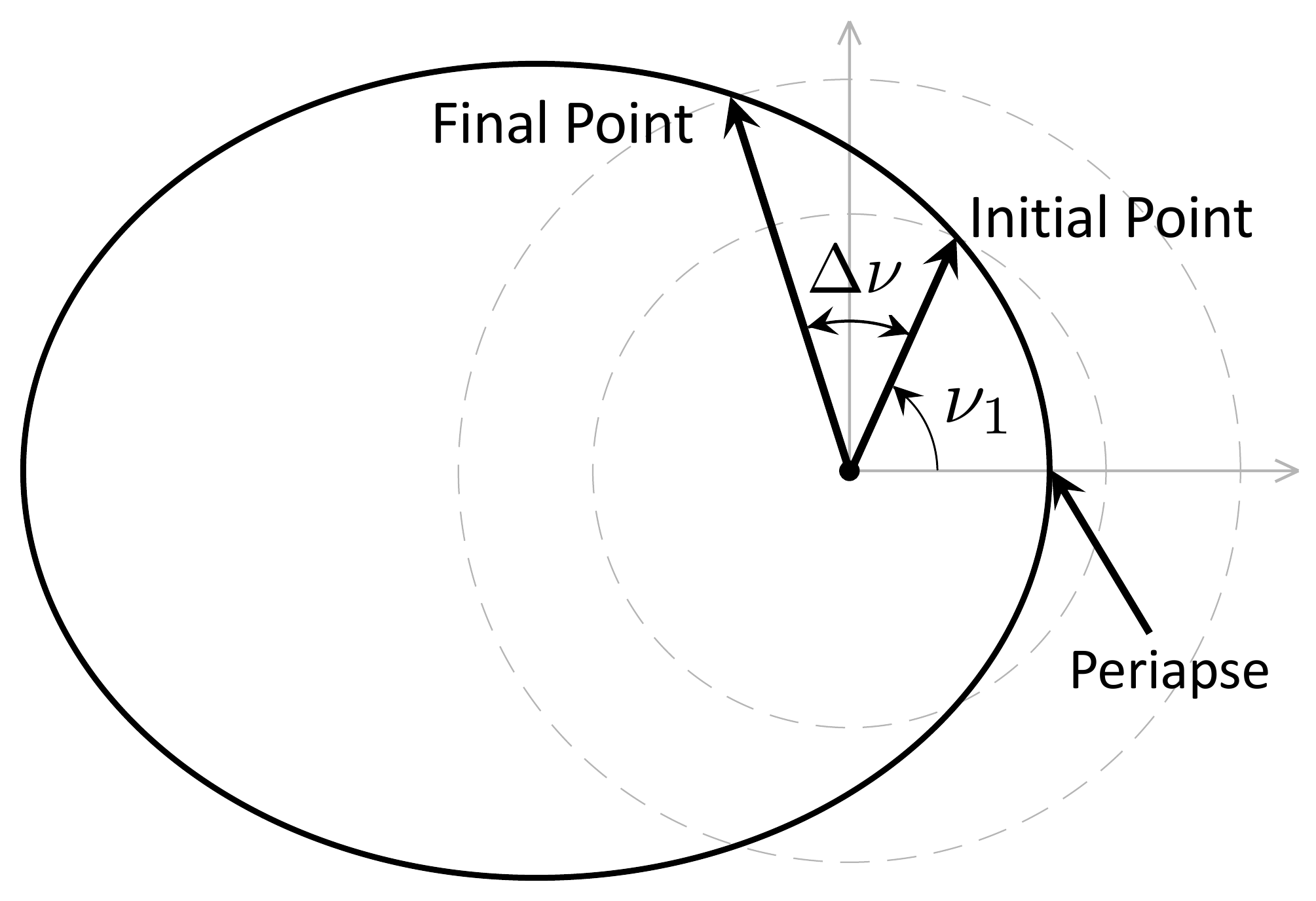}
\caption{Schematic showing the standard conic with $\nu_1$ and $\nu_2 = \nu_1 + \Delta \nu$.
}
\label{orbitfig}
\end{figure}

For two points with polar coordinates  $(R,\alpha)$ and $(R\gamma,\alpha+\phi)$  with
 $\gamma>1$,  the \textcolor{black}{first step of the} problem is to find the parameters of all conic arcs between them. 
The equation of a general conic has the form
\textcolor{black}{
\begin{equation}
  r(\theta)=\frac{p}{1+e \cos(\theta - \omega)}
\end{equation}
}
\textcolor{black}{where $\nu = \theta - \omega$} is the true anomaly \textcolor{black}{and $r$ is the radius.} 
A general conic is defined by its parameters  $(p,e,\omega)$. 
The scale parameter $p$ and the elliptic parameter, \textcolor{black}{or eccentricity,} $e$ are non-negative. 
A standard conic \textcolor{black}{as we define it here} has \textcolor{black}{the argument of periapse} $\omega=0$. 
Coordinate rotation by $\omega$ takes the conic with parameters $(p,e,0)$ to the conic with parameters $(p,e,\omega)$. 
Suppose a standard conic with parameters $(p,e,0)$  contains the points with polar coordinates $(1,\nu_1)$  and
 $(\gamma,\nu_1 +\Delta \nu)$
as shown in Figure \ref{orbitfig}.
\textcolor{black}{Here, the initial true anomaly $\nu_1$ is also referred to as the inside angle, and the change in the true anomaly $\Delta \nu$ is called the transfer angle.}
Then the rotated and rescaled conic with parameters
  $(Rp,e,\alpha-\nu_1)$ contains the points $ (R,\alpha)$ and 
   $(R\gamma,\alpha +\Delta \nu)$. 
We will rotate and rescale to produce transfer conics from standard conics.

\begin{figure}[ht!]
\centering
\includegraphics[width=0.5\textwidth]{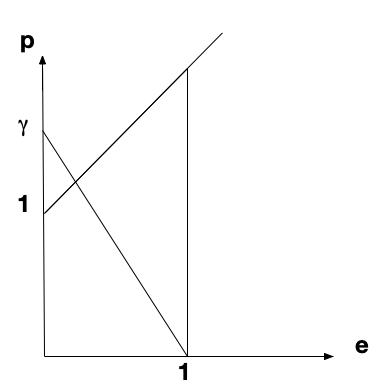}
\caption{
Allowed elliptic parameters
}
\label{fig1}
\end{figure}

\begin{figure}[ht!]
\centering
\includegraphics[width=0.8\textwidth]{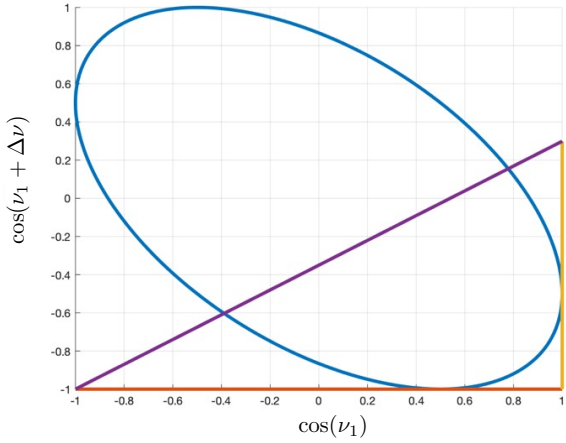}
\caption{
The set $Oval(\Delta \nu)$ and $T^*(\gamma)$ for $\gamma=1.524$  and $\Delta \nu=\pi/3$
}
\label{fig2}
\end{figure}

Given circles of radius $1$ and radius
 $\gamma>1$ centered at the origin of a plane, \textcolor{black}{we} first find all standard ellipses that intersect both circles. The minimum and maximum radii of an ellipse with parameters  $(p,e,0)$ are equal to $p/(1+e)$ and
  $p/(1-e)$, respectively. Necessary and sufficient conditions for intersections with the circles are
   $p/(1+e)\le 1 \le \gamma \le p/(1+e)$. 
The set of points  in the $(p,e)$ plane which satisfy these conditions forms a triangle
 $T(\gamma)$ 
 as shown in Figure \ref{fig1} for
  $\gamma = 1.524$. 
\textcolor{black}{
The value 1.524 was used because this is the roughly the distance of the orbit of Mars from the Sun in astronomical units.}
The triangle is bounded by the lines $e=1,p=1+e$, and $p=\gamma (1-e)$.
 For conics with $e\ge 1$ the condition for intersection is $p/(1+e)\le 1$.  
A standard conic with parameters $(p,e,0)$ generates two equations for its intersections with the two circles: 
\textcolor{black}{
\begin{equation}
\frac{p}{1+e \cos(\nu_1)}=1
\end{equation}
\begin{equation}
\frac{p}{1+e \cos(\nu_2)}=\gamma
\end{equation}
}
The cosines $(c_1,c_2)$ of the angles $(\nu_1,\nu_2)$ are found by solving the equations 
\begin{equation}
p=1+ec_1
\end{equation}
\begin{equation}
\frac{p}{\gamma}=1+ec_2
\end{equation}
The solution defines a map
\begin{equation}
f:T(\gamma)\to [-1,1] \times [-1,1]
\end{equation}
\begin{equation}
f(p,e)=(c_1,c_2)=((p-1)/e,(p-\gamma)/e\gamma)
\end{equation}			
The inverse of this map has the formula
\begin{equation}
f^{ -1 }(c_1,c_2)=(1+c_1(\gamma-1)/(c_1-\gamma c_2),(\gamma-1)/(c_1-\gamma c_2))
\end{equation}
The set $T^*(\gamma)=f(T(\gamma))$ is also a triangle contained in the square $[-1,1]\times [-1,1]$. The top boundary of $T^{*}(\gamma)$  is the image of the boundary of $T(\gamma)$ with $e=1$.  By using the formula for $f^{-1}$ one shows that this boundary is the line segment 
\begin{equation}	
c_2  =
\gamma^{-1}(c_1+1)-1 , \,
-1\le c_1 \le 1
.
\end{equation} 
The triangle $T^*(1.524)$ is shown in Figure \ref{fig2}. 
The cosine oval $Oval(\Delta \nu)$ shown in the plot is the parameterized curve 
$Oval(x) = (\cos(x), \cos(x+\Delta \nu) )$ for $0 \le x \le 2\pi$.
The intersection of the oval with the triangle will be used to solve for the
parameters of all standard ellipses that intersect the circles at points with
coordinates of the form $(1, \nu_1)$, and $(\gamma, \nu_1+\Delta \nu)$.
%

\begin{figure}[ht!]
\centering
\includegraphics[width=0.8\textwidth]{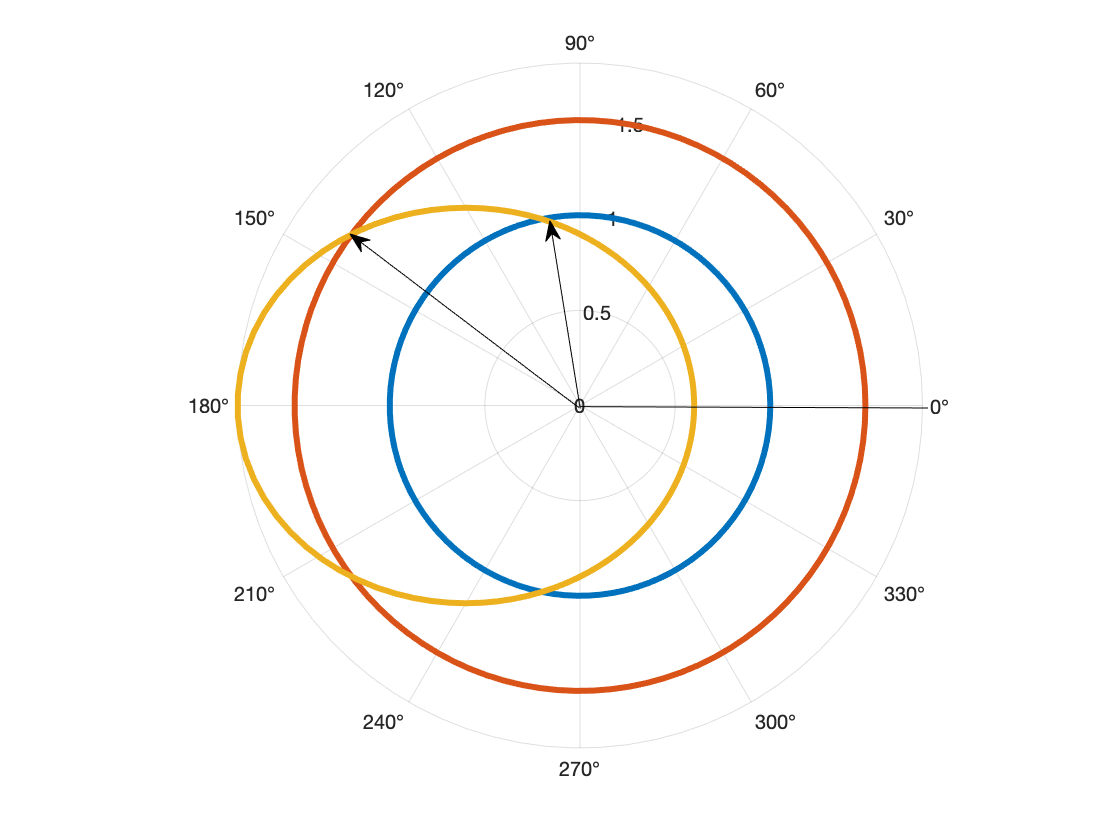}
\caption{
Circles of radius 1 and radius 1.5. Elliptic parameters $ p=.9, e=.5,\omega=0$. Angles  $\nu$ and $\Delta \nu$  are shown.
}
\label{fig3}
\end{figure}

\begin{figure}[ht!]
\centering
\includegraphics[width=0.8\textwidth]{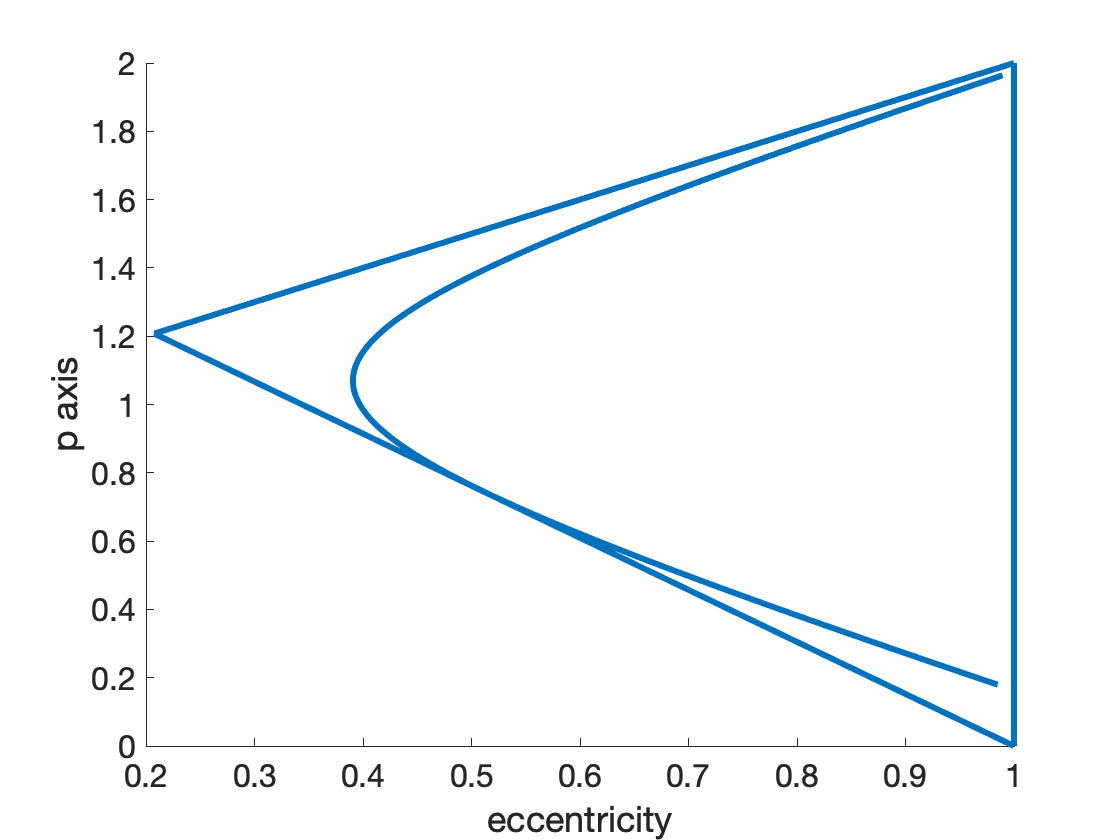}
\caption{
 The preimage of the set $T^*(\gamma)\cap Oval(\Delta \nu)$  viewed in the e-p plane.
}
\label{fig4}
\end{figure}

An example of a standard ellipse intersecting two circles is shown in Figure \ref{fig3}. 
For a given transfer angle  draw the oval and intersect this oval with the image
 $T^*(\gamma)$ of the map $f$. The intersection determines an interval of allowed inside angles
  $\nu_1(\Delta \nu,\gamma)\le \nu \le \nu_2(\Delta \nu,\gamma)$ . The intersection points of the cosine oval $Oval(\Delta \nu)$ with the upper boundary of the triangle $T^{*}(\gamma)$ are found by finding the solutions $\nu_1$ and $\nu_2$ of the equation
\textcolor{black}{
\begin{equation}
 \cos ( \nu_2 )
= \gamma^{-1}(\cos(\nu_1)+1)-1
\end{equation}
}
\textcolor{black}{where $\nu_2 = \nu_1 + \Delta \nu$.}
A typical intersection is shown in Figure \ref{fig2} for
 $\Delta \nu = \pi/3$. The preimage of the intersection $Oval(\Delta \nu)\cap T^{*}(\gamma )$ of the cosine oval as viewed in the $p$-$e$ plane is shown in Figure \ref{fig4}. This shows the elliptic parameters of the transfer ellipses with transfer angle $\Delta \nu$.
To compute the set of allowed elliptic parameters $f^{-1}[T^{*}(\gamma)\cap Oval(\Delta \nu)]$  one uses the map
\textcolor{black}{
\begin{equation}
g:[\nu_1(\Delta \nu,\gamma),\nu_2(\Delta \nu,\gamma)]\to R^2
\end{equation}
\begin{equation}
g(\nu_1)=f^{-1}(\cos(\nu_1),\cos(\nu_1+\Delta \nu))
\end{equation}
\begin{equation}
g(\nu_1)=(p(\gamma,\Delta \nu,\nu_1),e(\gamma,\Delta \nu,\nu_1))
\end{equation}
\begin{equation}
e(\gamma,\Delta \nu,\nu_1)=[\gamma-1]/[\cos(\nu_1)-\gamma \cos(\nu_1+\Delta \nu)]
\end{equation}
\begin{equation}
p(\gamma,\Delta \nu,\nu_1)=e(\gamma,\Delta \nu,\nu_1)\cos(\nu_1)+1
\end{equation}
}
The function $g$ determines the set of  allowed elliptic parameters of transfer ellipses as a function of the outer radius  and the transfer and inner angles. The function$f^{-1}$  applied to the set $Oval(\Delta \nu)$ also determines the set of all conic transfer parameters.				
To create elliptic arcs joining the points
 $(R_1,\alpha)$ and $(R_2,\alpha+\Delta \nu)$, choose $\gamma=R_2/R_1$.
Choose elliptic parameters$(p(\gamma,\Delta \nu,\nu_1),e(\gamma,\Delta \nu,\nu_1))$. These are the parameters of a standard ellipse that intersects the inner and outer circles with the transfer angle $\Delta \nu$ and inside angle
 $\nu_1$. Set $\omega=\alpha-\nu_1$. The angle  is the argument of the periapse of a rotated standard ellipse. Then the transfer ellipse between the points $(R_1,\alpha)$ and
 $(R_2,\alpha+\Delta \nu)$  is the general conic
 \begin{equation}
 r(\theta)=R_1p(\gamma,\Delta \nu,\nu_1)/[1+e(\gamma,\Delta \nu,\nu_1)\cos(\theta-\omega)]	\end{equation}
When the inside angle $\nu_1$ is unconstrained, the formula (16) produces parabolas and hyperbolas.

\section{Travel Times}
\textcolor{black}{
As a first step to determining the transfer orbit between two points for a specific time,
the transfer orbits for all times may be computed.
}
Consider travel times for a given gravitational parameter $\mu$, from an inner radius $R$ to an outer radius $\gamma R$  . The travel times depend on the values of 
$(\mu,R,\gamma,\Delta \nu,\nu_1)$.  Consider travel times for standard conics as a function of these variables. Use equations (14) and (15)  to produce parameters $p(\gamma,\Delta \nu,\nu_1)$, $e(\gamma,\Delta \nu,\nu_1)$ for a standard conic with the chosen transfer and inner angles. Then compute travel times using the following algorithm.
 
Algorithm: travel times as a function of $(\mu,R,\gamma,\Delta \nu,\nu_1)$.

1. Calculate $(e(\gamma,\Delta \nu,\nu_1),p(\gamma,\Delta \nu,\nu_1))=(e,p)$ using equations (14) and (15).

2. Set $P=Rp$  (rescale), $\sigma=\sqrt{\mu p}$ (angular momentum)

3. Integrate using the formula $r^2(\theta)d\theta=\sigma dt$

4. $T=\sigma^{-1} R^2p^2\int_{\nu_1}^{\nu_1+\Delta \nu} (1+ecos(\theta))^{-2} d\theta$
  (travel time) 
  
5.  For fixed parameters  $(\mu,R,\gamma,\Delta \nu)$   and for a given travel time $\tau$, solve the equation $ T(\mu,R,\gamma,\Delta \nu,\nu_1)=\tau$ for $\nu_1^*$ .
 
6. Calculate the parameters $(e(\gamma,\Delta \nu,\nu_1^*),p(\gamma,\Delta \nu,\nu_1^*))=(e^*,p^*)$ of the transfer ellipse.

\begin{figure}[ht!]
\centering
\includegraphics[width=0.8\textwidth]{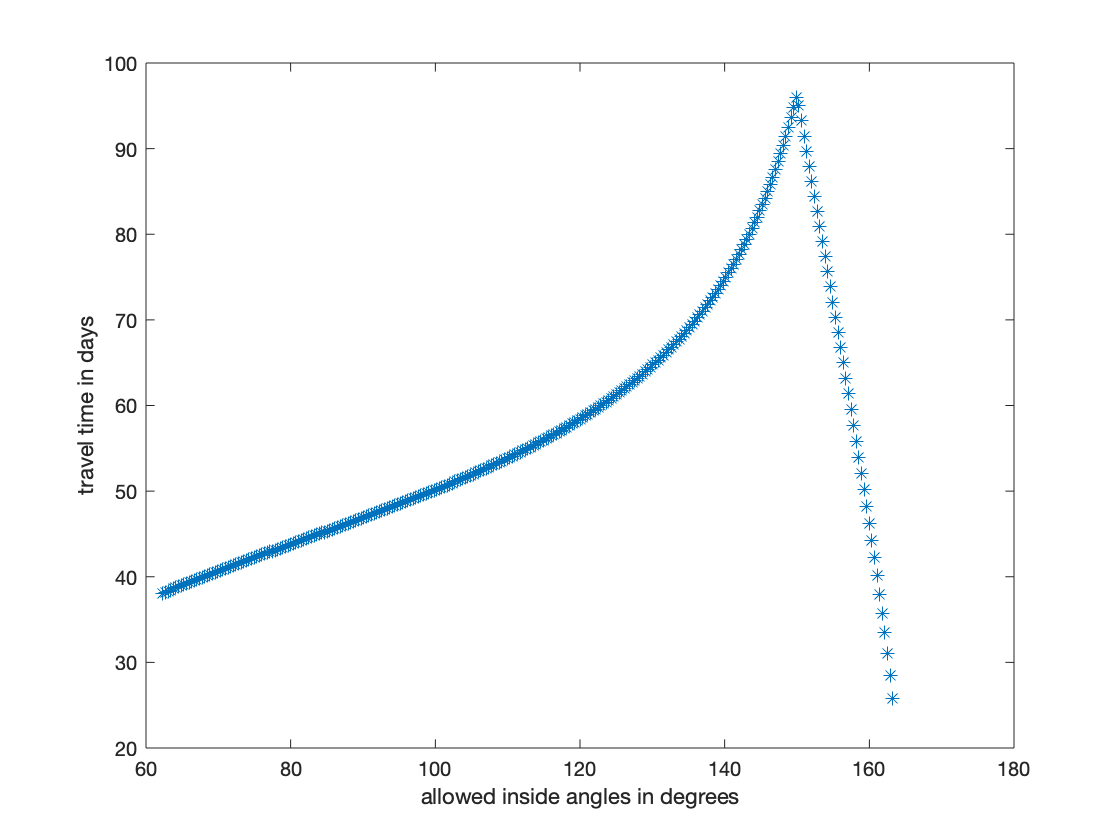}
\caption{ Travel times for allowed inside angles
 $\nu_1$ for $\gamma =1.524$ and $\Delta \nu =\pi /6$ 
}
\label{fig5}
\end{figure}

The time to travel from $(R,\,\alpha)$ to $(\gamma R,\alpha +\Delta \nu )$ on the transfer ellipse 
\begin{equation}
r(\theta)=Rp(\gamma,\Delta \nu,\nu_1 )/
[1+e(\gamma,\Delta \nu,\nu_1 )cos(\theta - \omega) ]
\end{equation}
with $\omega=\alpha - \nu_1$  is the same as the time to travel on the corresponding standard ellipse with $\omega = 0$ . Thus the travel time does not depend on $\omega$.  Steps 5 and 6 solve Lambert's problem.
Travel times for $\mu=1.327*10^{11}
\,km^3/s^2,\, \gamma=1.524, \,\Delta \nu_1=\pi/6$  are displayed in Figure \ref{fig5}  as a function of the admissible inside angles. 

\begin{figure}[ht!]
\centering
\includegraphics[width=0.8\textwidth]{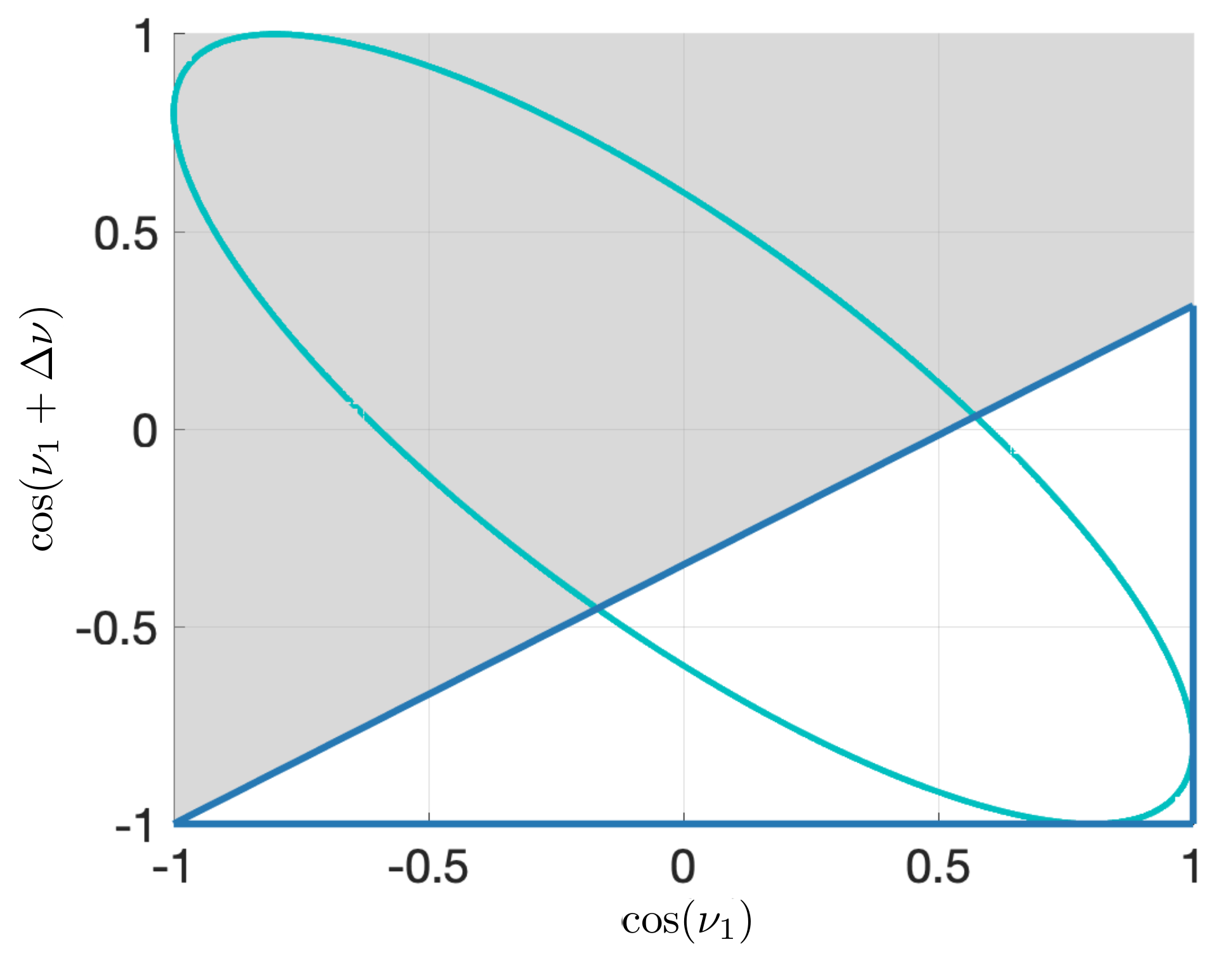}
\caption{The set $Oval(\Delta \nu)$ and $T^*(\gamma)$ for
$\gamma =1.524$ and $\Delta \nu = 143.2$ degrees. 
}
\label{newfig2}
\end{figure}

\clearpage
\textcolor{black}{
\section{Specific Travel Time Solution}
}
Once the possible transfer orbits have been computed as a function of the inside
angle $\nu_1$, the possible transfer orbits with a specific transfer time
may be computed.  A standard method for finding the zero of a function may then be used,
and some sample results were computed next using Matlab's \emph{fzero} function.
%
As an example, we consider the Mars 2020 transfer trajectory which launched on
July 30, 2020 and arrived on February 18, 2021. The angle between launch and
landing points is about \textcolor{black}{143.2} degrees, and the travel time
is about \textcolor{black}{203} days. The allowable range of inside angles
$(\nu_1, \nu_2)$ may first be computed by finding the intersections of the oval
with the triangle as shown
in Figure \ref{newfig2}.  The permissible values of $\cos(\nu_1)$ and $\cos(\nu_2)$ are given by the non-shaded region. 

\begin{figure}[ht!]
\centering
\includegraphics[width=0.8\textwidth]{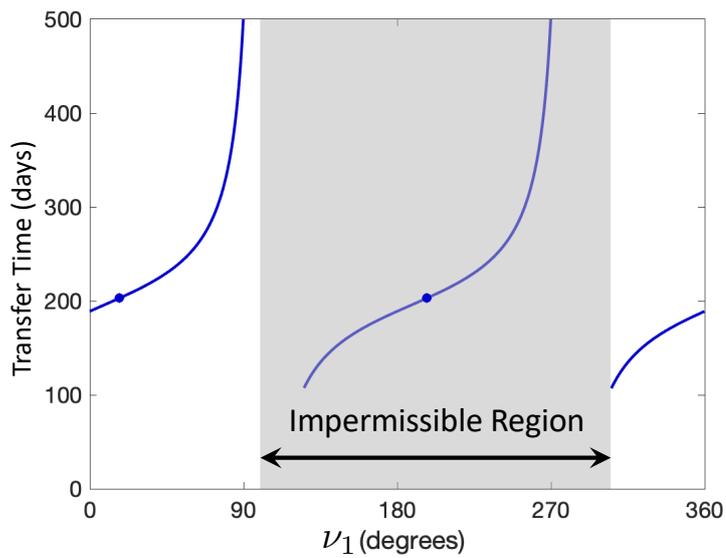}
\caption{
Travel time as a function of $\nu_1$ for the Mars 2020 trajectory.
}
\label{casesnew}
\end{figure}

\textcolor{black}{
By plotting the transfer time as a function of $\nu_1$, the different possible solutions between
the two desired points may now be clearly seen as shown in Figure \ref{casesnew}.  By using
different initial guesses for $\nu_1$, the algorithm can converge on possible different zeros 
along the curve.
}

\begin{figure}[ht!]
\centering
\subfigure[Dimensionless ellipse] {\label{subfig1}
\includegraphics[width=0.42\textwidth]{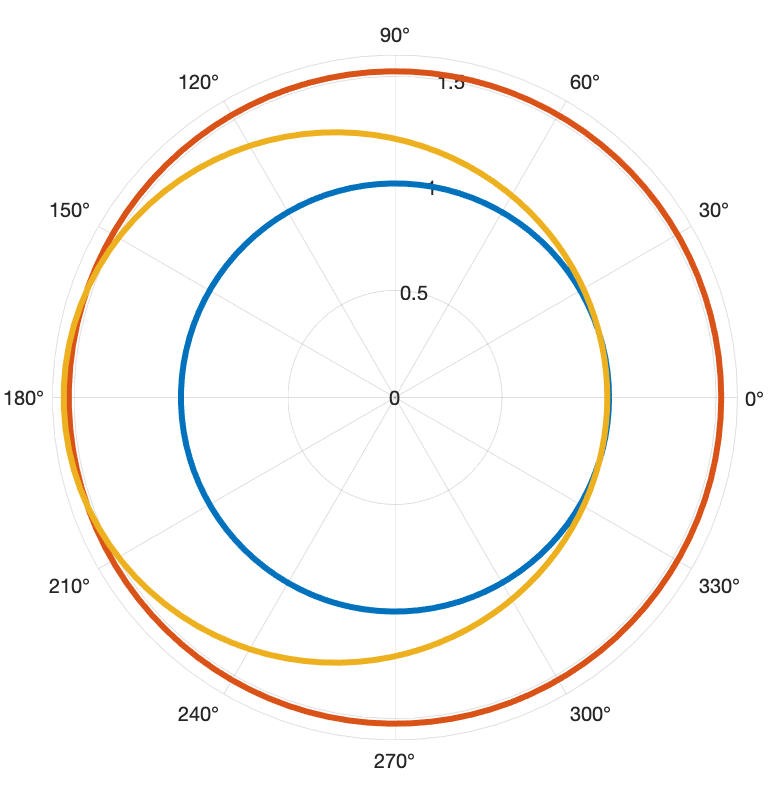}
}
\subfigure[Dimensional transfer] {\label{subfig2}
\includegraphics[width=0.50\textwidth]{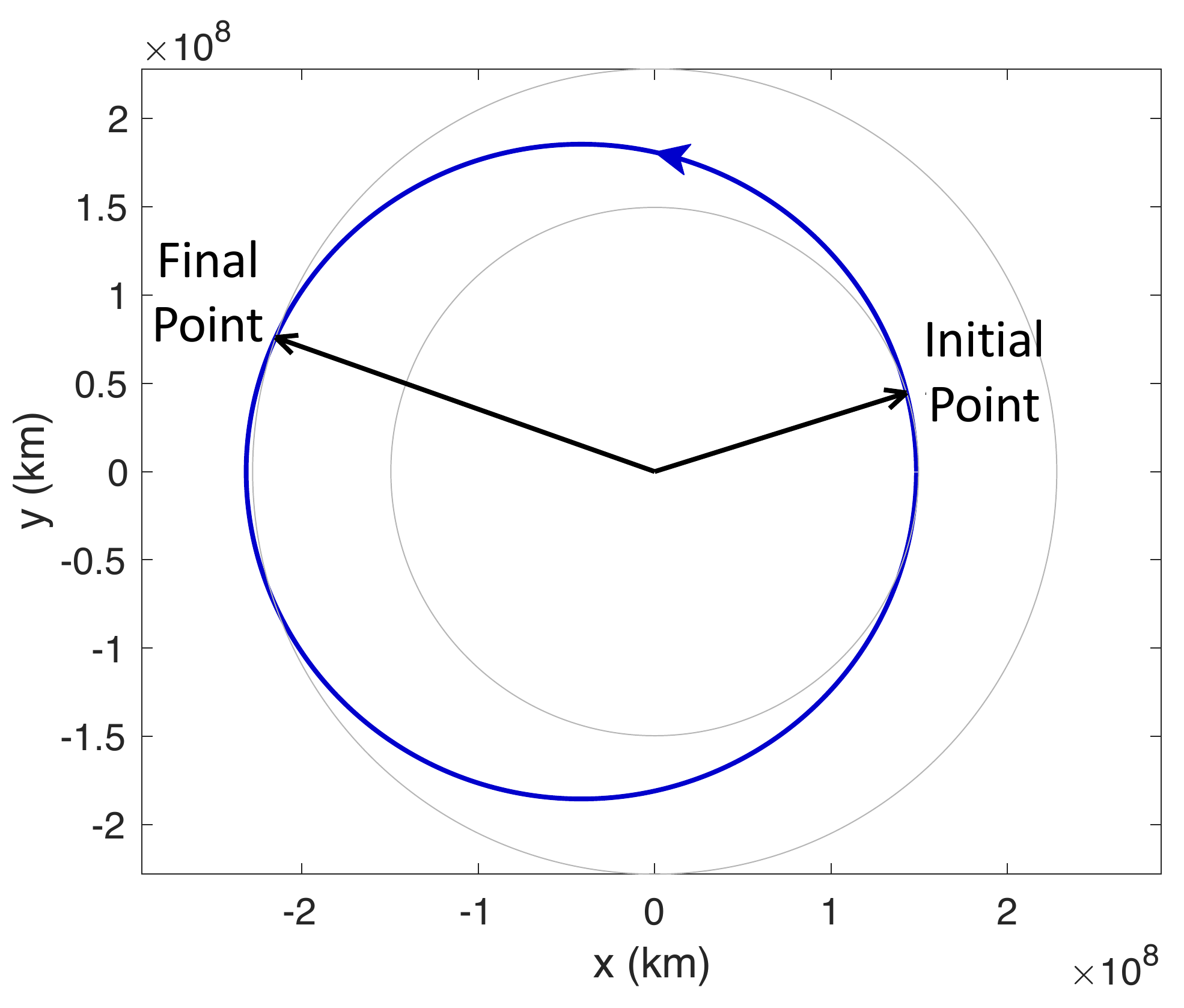}
}
\caption{
A transfer ellipse with elliptic parameters
$(e, p)$ = (0.21911558915832, 1.20917656075465)
is plotted along with circles of radius 1 and radius 1.524. The inside angle
 $\nu_1$ was approximately 0.302347076950009.
 }
\label{fig6}
\end{figure}

By limiting the search space to the allowable $\nu_1$ values and using the algorithm, 
the solution converged to $(\nu_1, e, p)$ = (0.302347076950009, 0.21911558915832, 1.20917656075465).
The
geometry of this ellipse is shown in Figure
\ref{fig6}.  \textcolor{black}{While these results were computed for the
dimensionless orbits in the plane, the actual transfer orbits may be computed
in three dimensions by simply scaling and rotating the resulting orbits in
combination with using states from the ephemeris.
}

\section{Conclusion}

A new elementary approach to Lambert's problem based on computing all conic
transfer orbits between two points is derived as a function of \textcolor{black}{the polar coordinate
of the points and the
initial true anomaly.} 
This method was
successfully used to compute these conic transfers across a range of possible
transfer times for the direct case using elliptic orbits.  By implementing a straightforward
zero finding method, it is possible to compute specific orbits with desired transfer times using
a standard conic.  This approach solves Lambert's problem once subsequent standard
rotations are performed to orient the orbit correctly in three-dimensional space.
The presented approach also allows the computation of various constraints on
the possible astrodynamic parameters for each transfer, which reduces the search space and
speeds up the computation of the desired solution.



\section{Acknowledgements}

Part of the research presented in this paper has been carried out at the Jet
Propulsion Laboratory, California Institute of Technology, under a contract
with the National Aeronautics and Space Administration. The authors would like
to thank James Meiss and David Sterling for valuable comments and suggestions.

%
%
%
%
%
%

\bibliographystyle{spbasic}
\bibliography{refs.bib}   

\begin{thebibliography}{7}
\providecommand{\natexlab}[1]{#1}
\providecommand{\url}[1]{{#1}}
\providecommand{\urlprefix}{URL }
\expandafter\ifx\csname urlstyle\endcsname\relax
  \providecommand{\doi}[1]{DOI~\discretionary{}{}{}#1}\else
  \providecommand{\doi}{DOI~\discretionary{}{}{}\begingroup
  \urlstyle{rm}\Url}\fi
\providecommand{\eprint}[2][]{\url{#2}}

\bibitem[{Bate et~al.(1971)Bate, Mueller, and White}]{Bate:1971}
Bate RR, Mueller DD, White JE (1971) Fundamentals of Astrodynamics. Dover
  Publications, Inc., New York

\bibitem[{Battin(1999)}]{Battin:1999}
Battin RH (1999) An Introduction to the Mathematics and Methods of
  Astrodynamics, {R}evised edn. AIAA Education Series

\bibitem[{Gooding(1990)}]{Gooding:1990}
Gooding RH (1990) A procedure for the solution of lambert's orbital
  boundary-value problem. Celestial Mechanics and Dynamical Astronomy
  48(2):145--165

\bibitem[{Pollard(1966)}]{Pollard:1966}
Pollard H (1966) Mathematical Introduction to Celestial Mechanics.
  Prentice-Hall Inc., Englewood Cliffs, New Jersey

\bibitem[{Prussing and Conway(2013)}]{Prussing:2013}
Prussing JE, Conway BA (2013) Orbital Mechanics, 2nd edn. Oxford University
  Press, New York

\bibitem[{Russell(2019)}]{Russell:2019}
Russell RP (2019) On the solution to every {L}ambert problem. Celestial
  Mechanics and Dynamical Astronomy pp 131--150

\bibitem[{Vallado(2013)}]{Vallado:2013}
Vallado DA (2013) Fundamentals of Astrodynamics and Applications, 4th edn.
  Space Technology Library, Microcosm Press

\end{thebibliography}

\appendix

\normalsize

\section{Appendix: Velocity Changes at Patch Points}

In this section the formulas for transfer conics and transfer times are applied to design patched conic orbits. To transfer from one arc to the next on a patched conic path, one must know the angle between the tangent lines to the arcs at the patch point. For a general conic with equation
\textcolor{black}{
\begin{equation}
r(\theta )=\frac{p}{1+e\cos(\theta-\omega )}
\end{equation}
}
The angle $\Phi$ of a unit tangent vector to the conic at a point $(r(\theta ),\theta )$ on the conic is required for this calculation. To find this angle one can calculate using complex variables. For this calculation the notation for eccentricity conflicts with exponential notation. Just for this calculation the symbol for eccentricity will be $\epsilon$.
\begin{equation}
q(\theta)=r(\theta)e^{i\theta},
\dot{q}(\theta)=
\dot{r}(\theta)e^{i\theta}+r(\theta)ie^{i\theta}=
s(\theta) e^{i\Phi}
\end{equation}
\begin{equation}
s(\theta)= \sqrt{\dot{r}(\theta )^2+(r(\theta )^2}
\end{equation}
\begin{equation}
\dot{r} (\theta )+r(\theta )i=s(\theta )e^{i(\Phi-\theta)}
\end{equation}
\begin{equation}
sin(\Delta \nu -\theta)=r(\theta) / s(\theta)
\end{equation}
\begin{equation}
\Phi=\theta + asin(r(\theta) /s(\theta))
\end{equation}
\begin{equation}
\dot{r}(\theta) = (\epsilon /p)r^2 (\theta)sin(\theta-\omega)
\end{equation}
\begin{equation}
\dot{r}(\theta)/r(\theta)=[1+\epsilon cos(\theta -\omega)]^{-1}\epsilon sin(\theta-\omega)
\end{equation}
\begin{equation}
r(\theta) /s(\theta)=1/\sqrt{\epsilon^2sin^2(\theta -\omega) /
{[1+\epsilon cos(\theta -\omega) ]} ^2 +1}
\end{equation}
Now replace $\epsilon$ by $e$ as the symbol for eccentricity. The result is
\begin{equation}
\Phi = \theta + asin(1/\sqrt{\epsilon^2sin^2(\theta -\omega) /
r^2(\theta)+1}\,)
\end{equation}
The energy of the orbit in terms of the elliptic parameters $(p,\, e,\, \omega)$ is
$E=\mu(e^2-1)/2p$.
For an elliptic orbit of the Kepler problem
 $\ddot{q}=-\mu r^{_3}\, \,with\, r=|q|$
The Cartesian velocity $v(r,\theta)=\dot{q}$  is the velocity at the point $(r,\theta)$ on the conic and the speed is
  \begin{equation}s(r,\theta ) =|\dot{q}|=\sqrt{2(E+\mu r^{-1})}=
\sqrt{2\mu} \sqrt{[(e^2 -1)/p] +r^{-1}}
\end{equation}
 The velocity in polar coordinates is
  $(s(r,\mu,e,p),\Phi(r,\theta,e,p,\omega))$.
To transfer from an incoming elliptic orbit to an outgoing elliptic orbit at a common patch point
 $(r,\theta)$ the velocity change needed is the difference between the outgoing and incoming velocities at this point. These velocities are determined for polar coordinates by the above formulas.

\begin{figure}[ht!]
\centering
\includegraphics[width=0.8\textwidth]{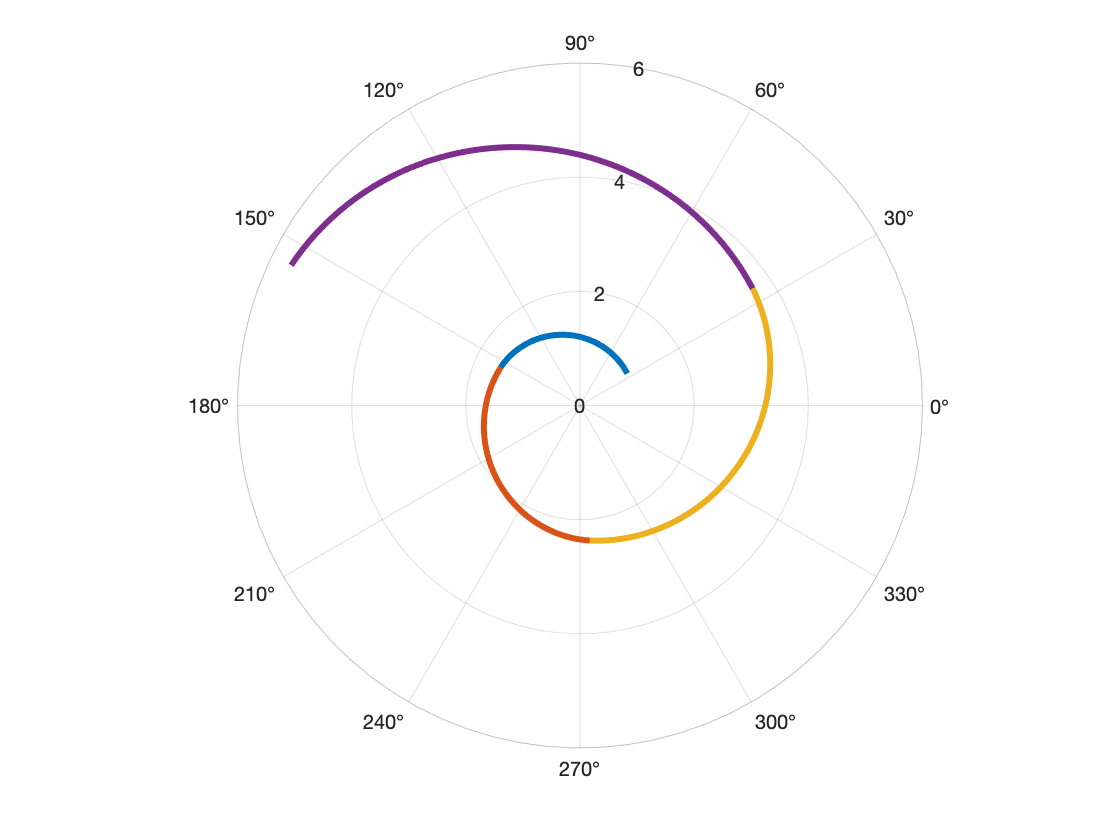}
\caption{ Four conic arcs with  $\gamma=1.524,\,\Delta \nu=2\pi/3 \,\,p_0=1.202,\,e_0=.2385$
 }
\label{fig7}
\end{figure}

To patch elliptic arcs in an outward spiral one can go back to geometry and choose conic arcs between circles of radii $1,\,\gamma,\gamma^2,...$ with a fixed transfer angle $\Delta \nu$ between patch points and choose allowed elliptic parameters
 $(1\, ,e_0 ,\omega_0)$. Then choose an initial scale parameter  and generate a sequence of elliptic parameters $p_1 = \gamma p_0 , \, e_1=e_0 , \omega_1= \omega_0 + \Delta \nu,\,...$
Elliptic arcs, travel times, and velocity changes can now be computed from this data. A plot for the choice $\gamma=1.524,\,\Delta \nu=2\pi/3,\,p_0=1.202,\,e_0=.2385$ is shown in Figure \ref{fig7}.




\end{document}